\documentclass[letter]{aa}

\usepackage{graphicx}
\usepackage{txfonts}
\begin{document}

   \title{Gravitational collapse of the OMC-1 region \fnmsep\thanks{Based on observations carried out with the IRAM 30m Telescope. IRAM is supported by INSU/CNRS (France), MPG (Germany) and IGN (Spain).}}


   \author{A. Hacar
          \inst{1,2}
          \and J. Alves\inst{1}
          \and
          M. Tafalla\inst{3} \and J. R. Goicoechea\inst{4}
          }
         \institute{Institute for Astrophysics, University of Vienna,
              T\"urkenschanzstrasse 17, A-1180 Vienna, Austria
         \and
         	   Leiden Observatory, Leiden University, P.O. Box 9513, 2300-RA Leiden, The Netherlands\\
         	   \email{hacar@strw.leidenuniv.nl}
         \and
             Observatorio Astronomico Nacional (IGN), C/ Alfonso XII, 3, E-28014, Madrid, Spain
	    \and
             Instituto de Ciencias de Materiales de Madrid (CSIC), C/ Sor Juana Ines de la Cruz 3, E-28049 Cantoblanco, Madrid, Spain
             }

\abstract{
We have investigated the global dynamical state of the Integral Shaped Filament in the Orion A cloud using new N$_2$H$^+$ (1-0) large-scale, IRAM30m observations. Our analysis of its internal gas dynamics reveals the presence of accelerated motions towards the Orion Nebula Cluster, showing a characteristic blue-shifted profile centred at the position of the OMC-1 South region. 
The properties of these observed gas motions (profile, extension, and magnitude) are consistent with the expected accelerations for the gravitational collapse of the OMC-1 region and explain both the physical and kinematic structure of this cloud. 
} 


   \maketitle
%

\section{Introduction}

Observations suggest the gravitational collapse of the gas within young clusters as a key ingredient determining the dynamical properties of these systems. 
In a recent work, \citet{KIR13} investigated the presence of gravitationally-induced motions in the Serpens-South region.  
From the analysis of the N$_2$H$^+$ emission as unambiguous tracer of the dense, UV-shielded molecular gas \citep{PET16}, Kirk et al identified a longitudinal velocity gradient as part of the filamentary accretion inflow toward this proto-cluster. \citet{PER14} attributed the detection of similar linear gradients to the homologous collapse along filaments feeding the central clump of the SDC13 infrared dark cloud. Increasing evidence indicates that these gravitationally dominated accretion flows might also be involved in the formation of massive stars
	\citep[e.g.,][]{GAL10,PER13}.

If generated by a free-fall collapse, gravity is expected to produce a characteristic signature on the gas velocity structure. Due to the radial dependency of the gravitational attraction, models of clouds in gravitational collapse \citep[e.g.,][]{GOM14} predict these motions to be accelerated towards the center of the potential well. Despite these expectations, such gravitationally driven accelerations remain poorly characterized in observations.

In this letter, we investigate the dynamical state of the dense gas along the Integral Shaped Filament (ISF) in Orion \citep{JOH99}. As illustrated in Fig.~\ref{fig:map}, the ISF harbors the Orion Nebula Cluster (ONC), the most active stellar cluster in the solar neighborhood and the nearest site for high-mass star formation, including the Trapezium stars and the Orion KL region \citep[see ][for a review]{ODE01}. Still partially embedded in the ISF, the ONC is found in association with large amounts of dense material within OMC-1 cloud, a region widely investigated in the past at millimeter wavelengths \citep[e.g.,][]{MPI90,ROD92,WIS98}. 
Our results demonstrate the presence of accelerated gas motions along the OMC-1 cloud consistent with the expected gravitational collapse of this region.

\section{IRAM30m observations}

\begin{figure*}
   \centering
   \includegraphics[width=0.9\linewidth]{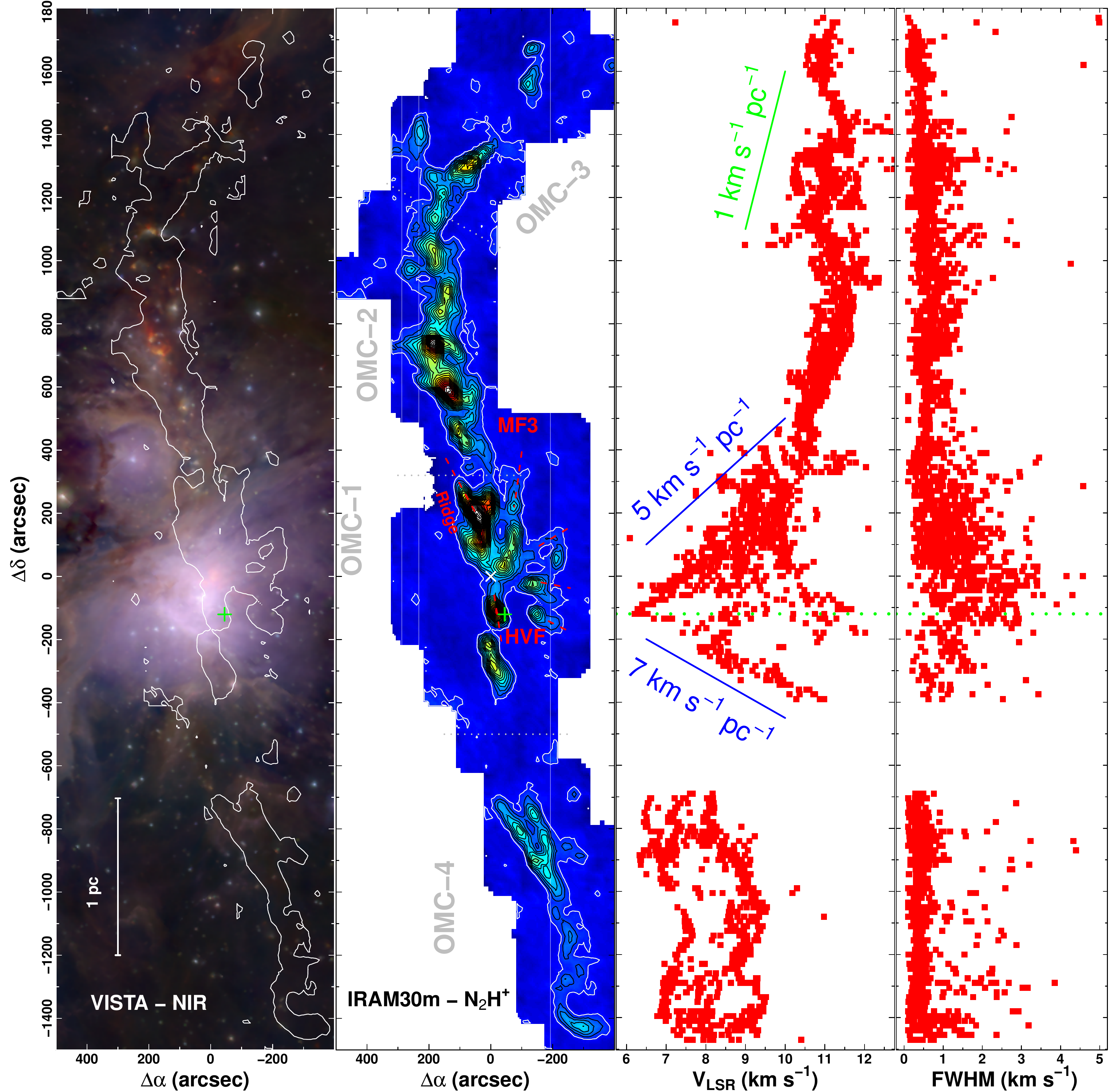}
      \caption{From left to right: (a) VISTA-NIR image of the ISF \citep{MEI16}; (b) IRAM30m integrated N$_2$H$^+$ (1-0) emission; (c)  N$_2$H$^+$ line velocity centroid (V$_{LSR}$) and (d)  N$_2$H$^+$ Full-Width-Half-Maximum (FWHM) as a function of declination. Map offsets are referred, in radio projection, to the position of the Orion BN source $(\alpha,\delta)_{J2000} = (05^h35^m14^s.2,-05º22'21'')$ (white cross).
      N$_2$H$^+$ contours are equally spaced every 2~K~km~s$^{-1}$. For reference, the first N$_2$H$^+$ contour is superposed to the IR image. The location of the most prominent molecular fingers \citep[e.g., the OMC-1 ridge, MF3, and HVF;][]{ROD92} are indicated by red lines in the integrated intensity map. The magnitude of representative gradients with 1 (green) plus 5, and 7 (blue) km~s$^{-1}$~pc$^{-1}$ are indicated in the velocity plot. The position of the OMC-1 South clump with the most blue-shifted velocity is indicated in both maps (green cross) and kinematic plots (green dashed line).}
             \label{fig:map}
   \end{figure*}

We observed the entire ISF in November 2013 using the IRAM30m telescope.
We observed this region at the frequency of the N$_2$H$^+$ (1-0) line \citep[93173.764 MHz,][]{PAG09} using the EMIR receiver connected to the VESPA spectrometer achieving an effective spectral resolution of 0.06~km~s$^{-1}$.  The observations were carried out in Position-Switching mode using an OFF position at approximately 1 degree from the center of our maps with no detected emission above 0.02~K. Calibrations were carried out every 15~min as well as pointing and focus every 1.5-2.0~hours. The reduction strategy, baseline subtraction, and main beam calibration follow the reduction scheme presented by \citet{HAC17}. Our final map covers a total area of $\sim$~420 arcmin$^2$ obtained from a mosaic with different tiles of 200~$\times$~200 and 100~$\times$~100 arcsec$^2$ each.  A final Nyquist sampled set of 7030 spectra are obtained after the convolution of the original dataset into a final resolution of 30'', showing a typical rms of 0.15~K. All the data presented in this work will be released after their combination with existing interferometric observations (Hacar et al, in prep.). 

Figure~\ref{fig:map}b illustrates the total integrated intensity map of the N$_2$H$^+$ (1-0) emission  within this region. 
The coverage of our maps corresponds to a $\sim$~7-pc long, north-south longitudinal cut along the entire ISF, including the ONC \citep[e.g.,][]{ODE01}, the Orion BN/KL region, the Orion Bar, and the Orion HII region around the Trapezium \citep[e.g.,][]{GOI15} (see also Fig.\ref{fig:map}a). Tracing the gas at densities of n(H$_2$)~$>$~10$^{4.5}$~cm$^{-3}$ \citep[e.g.,][]{CAS02}, the N$_2$H$^+$ emission reveals the intricate distribution of dense gas along the main spine of the ISF. Several branches and substructures are identified in regions like OMC-2/3, in agreement with previous results \citep[e.g.,][]{LI13}. Additional features, such as the molecular fingers \citep{MPI90,ROD92} and the OMC-1 South region \citep{MEZ90} are clearly distinguished by their bright N$_2$H$^+$ emission. Disconnected from the rest of the cloud, the emission of this molecule is recovered towards the south, coincident with the OMC-4 region. The distribution of the N$_2$H$^+$ (1-0) emission in our maps is  in good agreement with previous studies in the same region \citep{TAT08}. Compared to these observations, however, our data is more sensitive and is better spectrally and spatially resolved by a factor of 1.5-2.  

\section{Accelerated gas motions towards the ONC}\label{sec:acceleration}

We investigated the internal gas motions in the ISF from the detailed analysis of the N$_2$H$^+$ (1-0) emission profiles. This information is obtained from the hyperfine fit of the individual spectra following the same multi-line fitting strategy presented by \citet{HAC17}. This fitting procedure provides a full description of the gas emission in terms of the mean gas velocity along the line-of-sight (i.e., velocity centroids), the FWHM,  line emission properties (i.e., intensity and opacity), and line multiplicity (or number of components) at each position in our map. 

Using the above fitting procedure, we recover a total of 4207 components with S/N~$\ge$~3. At the angular resolution of our single-dish observations, a large fraction of the spectra present a single line component. Nevertheless, extended areas exhibiting well-separated, double peaked spectra are found in regions like the OMC-1 ridge and the OMC-4 cloud. Due to the limited beam size of our IRAM30m observations, in this work we restrict our study to the gas motions within the ISF at scales $\gtrsim$~0.25~pc. The internal physical and kinematic substructure of this region will be investigated in higher detail in a subsequent paper combining both the current single-dish observations and a new set of ALMA interferometric data (Hacar et al, in prep.). 

Figure~\ref{fig:map}c shows the mean velocity of the gas along the line-of-sight (V$_{LSR}$) as a function of declination for all the detected components in our maps (red points).
Previous studies using observations of different CO isotopologs have pointed out the existence of a global north-south velocity gradient of $\lesssim$1~km~s$^{-1}$~pc$^{-1}$ in ISF \citep[green line;][]{BAL87} continuing along the entire Orion A cloud \citep{HAC16}. A similar trend is observed in the dense gas detected in N$_2$H$^+$ at large scales, with velocities ranging between V$_{LSR}\sim$~11.5~km~s$^{-1}$ in OMC-3 and $\sim$~8~km~s$^{-1}$ in OMC-4.

In addition to the above global gradient, Fig.~\ref{fig:map}c reveals a rapid change of the gas velocities in the proximity of the ONC region. The observed N$_2$H$^+$ emission exhibits a strong blue-shifted drift towards the OMC-1 cloud. Similar gas motions have been suggested from the analysis of the gas kinematics along the main axis of the OMC-1 gas molecular fingers \citep{ROD92,WIS98}. 

When observed at large scales, the velocity structure within the OMC-1 cloud describes a characteristic V-shape centered around the OMC-1 South clump. In the proximity of the ONC, we identify gradient values up to 5-7~km~s$^{-1}$~pc$^{-1}$ (blue lines), that is, between 5-10 times larger than those values measured in other regions of the same complex, such as OMC-2 or OMC-3. The systematic nature of these motions indicates that the gas is accelerated in the directions of the OMC-1 region.

\section{Radial dependency of the gas motions within the OMC-1 region}\label{sec:radial}

\begin{figure}
   \centering
   \includegraphics[width=1.0\linewidth]{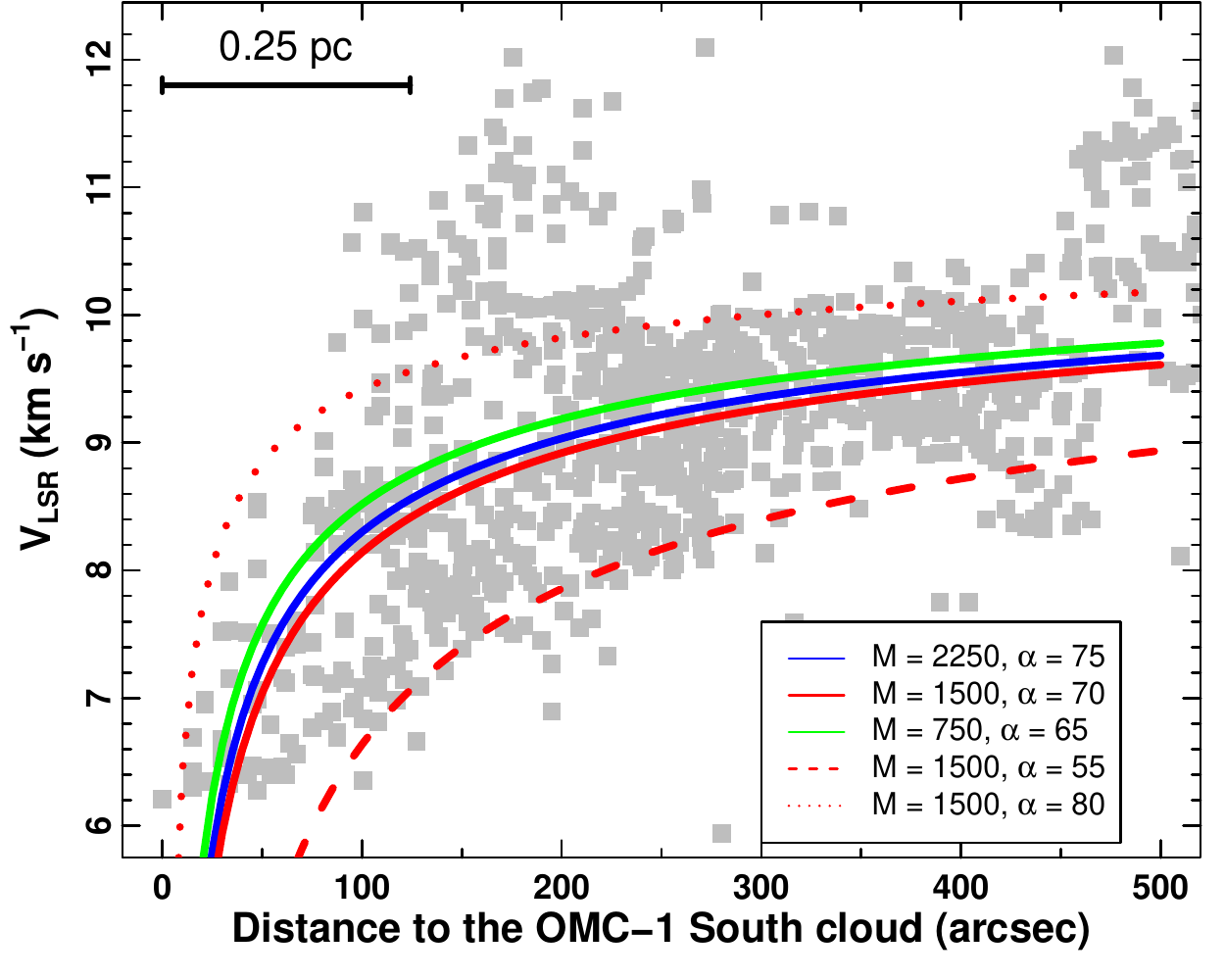}
      \caption{Gas velocity structure as a function of the distance to the OMC-1 South cloud for all the gas components detected with S/N~$\ge$~3 (gray squares). Different lines describe the expected velocity profile for a free-falling particle in a series of potential wells with different masses M observed at different angles $\alpha$ following Eq.~\ref{eq:infall} (see parameters in the lower right corner; masses are in M$_\sun$ and angles are in degrees). }
             \label{fig:collapse}
   \end{figure}

The blue-shifted velocity feature identified in Fig.~\ref{fig:map} appears to present a radial dependence respect to the position of the OMC-1 South clump. 
We have highlighted this behavior in Fig.~\ref{fig:collapse} by plotting the gas velocity structure as a function of the distance to coordinates (x,y)=(45'',-120'') in our maps, selected as the position of most blue-shifted line centroid within the OMC-1 region. As deduced from this plot, the reported velocity gradient extends radially across the entire OMC-1 region and up to distances of $\sim$~400 arcsec from this central position, or $\sim$~0.8~pc at the distance  of Orion \citep[D=414~pc][]{MEN07}.

Although less systematic than the velocity shifts detected in the line centroids, a similar radial dependency is observed between the derived N$_2$H$^+$ FWHM as a function of the distance to the OMC-1 South region (see Fig.~\ref{fig:map}d). In relative terms, the observed FWHM rises from approximately 0.5-1.0~km~s$^{-1}$ in regions like OMC-2 or OMC-4 to up to 2.5-3.0~km~s$^{-1}$ in the area surround the OMC-1 South cloud. 
Part of these effects can be explained by thermal broadening effects in the proximity of the Trapezium stars produced by the increasing gas temperatures measured in this region \citep{WIS98}. In addition to this, the parallel evolution of the observed FWHM with the gas velocity gradients suggests that part of the reported enhanced FWHM values could be generated by the increasing contribution of the large-scale motions on the velocity field sampled along the line-of-sight (l.o.s.).

Interestingly, the direction of the gas velocity gradients within the OMC-1 region closely corresponds with the direction of the well characterized molecular fingers \citep{ROD92,WIS98}. 
When isolated by their integrated emission, the gas velocity field along each of these structures describes a continuous velocity track as a function of radius with remarkably small deviations from their large-scale behavior \citep[e.g., MF3 or HVF in][; see also Fig.~\ref{fig:map}~b]{ROD92}. With perhaps the exception of the more complex OMC-1 ridge, the systematic organization of the observed physical and kinematic structrure of the gas at large scales might explain the origin of the molecular fingers within the OMC-1 region. 

\section{Gravitational collapse of OMC-1 region}

The characterization of the line centroid of molecular lines and, in particular, their Doppler velocities provide information on the bulk motions of the gas along the l.o.s..
Due to the lack of information about the real  3D-space, however, the physical interpretation of the relative motions between superposed structures usually lead to ambiguous conclusions. Depending on their relative orientation, the observational detection of velocity gradients and/or velocity differences can be equally interpreted as both converging or diverging motions \citep[e.g., see Fig.~12 in][]{HEN14}. 

Similar to our results, the presence of a blue-shifted, V-shape velocity structure of the dense gas associated ONC was first reported by \citet{ROD92} from the study of the HC$_3$N emission within the OMC-1 cloud. From the comparison with lower density tracers, Rodriguez et al suggested that the gas molecular fingers are produced by the interaction of the ionized and neutral gas. In particular, these authors interpret the abrupt variation of the gas velocity as the direct impact of the shock-front with the pre-existing dense gas in the area surrounding of the Trapezium region, producing a pillar-like structure showing a velocity shift respect to the outermost gas \citep[see also][]{MUR90}.

Different observables seem to be in tension with this sweep-up scenario. 
First, this pillar-like configuration should favor the formation of cometary shapes with characteristic head-tail enhancements, similar to those clumpy regions exposed to stellar winds, such as the Horsehead Nebula \citep[e.g.,][]{HIL05}, not observed in any previous molecular dataset within this region. 
More importantly, it is unclear how this formation mechanism would mimic the velocity structure and continuity of the cloud at large scales in regions not affected by these winds (e.g., OMC-2/3; see Fig.~\ref{fig:map}~c). Although some of the observed molecular fingers could indeed be shaped by the strong stellar feedback within the ONC region \citep[e.g., OMC-1 South,][]{ODE01}, 
winds alone seem an unlikely mechanism for the formation of these objects.

Alternatively, we suggest that the velocity gradient observed towards the OMC-1 region actually corresponds with dense gas material moving in the direction of the observer and towards the position of the ONC. The low extinction values measured toward the Trapezium region demonstrate that most of the dense gas detected along the OMC-1 ridge is located behind this stellar cluster \citep[e.g.,][]{ODE01}. This particular configuration allows us to break the degeneracy: red-shifted gas velocities would then be interpreted as outflowing motions along the l.o.s. and moving away from the stars. On the other hand, the observed blue-shifted velocities would be indicative of infalling material towards the cluster center. 

As a unique characteristic of gravitationally generated motions, a rapid acceleration is expected in the proximity of the potential well. It is easy to prove that for a point-like mass in free-fall, the observed l.o.s. velocity V$_{LSR}$ at a given impact parameter {\it p} can be described by the following relation:
\begin{equation}\label{eq:infall}
	V_{LSR}(p)=V_{sys,0}+V_{infall}(p) \cdot cos(\alpha)
\end{equation}
where $V_{sys}$ is the initial systemic velocity and V$_{infall}(p)=-\sqrt{\frac{2GM}{R}}=-\sqrt{\frac{2GM}{p\ /sin \alpha}}$ the infall velocity in a potential of a mass M as a function of the distance R to its center, respectively, with $\alpha$ describing the orientation angle of these infalling motions with respect to the l.o.s..
In such configuration, the observed velocities are expected to present an approximately  $\sim p^{-1/2}$ functional dependence with respect to the center of the potential well, modulated by their viewing angle $\alpha$.

In order to describe the global motions within the OMC-1 region, in Fig.~\ref{fig:collapse} we display the expected velocity profiles for particles in free-fall in three potential wells with masses of 2250 (blue), 1500 (red), and 750 (green) M$_\sun$, all of them presenting a systemic velocity V$_{sys,0}=$~10.8~km~s$^{-1}$ similar to the observed gas velocities at the northern end of the OMC-1 region (see Fig.~\ref{fig:map}). In addition to this, we explore different projection angles $\alpha$ leading to reasonable fits of the observed gas velocities. In all cases, the kinematic signature of the gravitational infall is assumed to be blue-shifted with respect to the systemic velocity given the physical 3D configuration of this cloud (see above). 

As seen in Fig.~\ref{fig:collapse}, our toy model closely reproduces most of the bulk motions observed along the OMC-1 region. We find remarkably good agreement between our fitting values and the measured masses for the ONC, estimated in $\sim$~1800~M$_\sun$ \citep[with a r$_{0}\sim$~0.2~pc,][]{HIL98}, assumed as the dominating component of the gravitational potential over the $\lesssim$~600~M$_\sun$ of gas detected within the same region \citep{GOL97}. 
Moreover, our models predict the molecular fingers to be oriented close to the plane of the sky (i.e., $\alpha \sim$~55-80 deg), explaining their long projected dimensions observed at mm-wavelengths \citep[e.g.,][]{WIS98}. 
Deviations from this global behaviour can be explained by the intrinsically more complex structure of the OMC-1 region and the distinct orientation of each individual molecular finger compared to our oversimplified model. 
Still, the good agreement of the above results indicates that most of the accelerated motions identified within the OMC-1 region could actually correspond to the kinematic signature generated by its gravitational collapse. 

\section{Discussion and conclusions}

In this letter we have investigated the global velocity structure of the ISF in Orion using a new series of IRAM30m N$_2$H$^+$ (1-0) observations. Compared to the smooth velocity gradients dominating the global velocity structure of this cloud, our large-scale maps demonstrate the presence of accelerated motions towards the OMC-1 region with gradients up to 7~km~s$^{-1}$~pc$^{-1}$. When displayed in a Position-Velocity diagram, these motions exhibit a characteristic blue-shifted, V-shape centered at the position of the OMC-1 South.

The direction of the maximum gradients observed along the OMC-1 region
seem to coincide with the orientation of the so-called molecular fingers. 
Although initially interpreted as ejected material, 
the large-scale continuity of these motions and the well characterized physical structure of this region  suggest that the observed V-shaped profile may correspond to an accelerated material inflowing towards the ONC. The bulk of these motions, as well as their radial dependency, can be reproduced by a simplified infall model, whose physical properties mimic the expected mass, orientation, and 3D structure of the ONC region. We conclude that both the mass distribution and gas kinematic within the OMC-1 region might be dominated by the gravitational collapse of this cloud. 

If confirmed, the proposed collapse of the OMC-1 region could potentially provide a significant mass accretion flow onto the ONC cluster. For a filamentary inflow, the mass accretion rate can be described by $\dot{M}=\pi r^2 \left< n(H_2)\right> \times V_{infall}(R)$ \citep[e.g., see][]{KIR13}. Falling from a distance R=0.8~pc (see Sect.~\ref{sec:radial}), and with a typical radius of r~$\sim$~0.05~pc and an average density $\left< n(H_2)\right>\sim$~10$^5$~cm$^{-3}$, the mass inflow per molecular finger is estimated in $\dot{M}_i\sim$~55~M$_\odot$~Myr$^{-1}$. The 7 molecular fingers detected in OMC-1 \citep{ROD92} would then contribute with a total $\dot{M}(total)\sim385$~M$_\odot$~Myr$^{-1}$. 

Our results in the OMC-1 region provide new evidence indicating gravity as a key ingredient for the formation of massive clusters. Interestingly, the gravitational collapse of the OMC-1 region appears to continue after the formation of the main cluster core suggesting that the cluster assembling phase might expand on timescales of several Myr. In the particular case of the ONC, the still on-going collapse of the OMC-1 region could potentially explain the elongated, and still not relaxed, nature of this cluster. Moreover, the continuous inflow of fresh material contained in these star-forming molecular fingers brings a new perspective on the stellar mass segregation reported in previous studies \citep{HIL98}.

\begin{acknowledgements}
      This work is part of the research programme VENI with project number 639.041.644, which is partly financed by the Netherlands Organisation for Scientific Research (NWO). MT and AH thank the Spanish MINECO for support under grant
AYA2016-79006-P. JRG and MT thank the Spanish MINECO for support under grant AYA2012-32032.
\end{acknowledgements}

%
%


\end{document}